# Flexible complementary logic circuit built from two identical organic electrochemical transistors

L. Travaglini[1,2], A. P. Micolich[3],* C. Cazorla[1], E. Zeglio[1,2,4], A. Lauto[5], D. Mawad[1,2,5]*

[1]School of Materials Science and Engineering, UNSW Sydney, Sydney, New South Wales 2052, Australia

[2]Centre for Advanced Macromolecular Design, UNSW Sydney, Sydney, New South Wales 2052, Australia

[3]School of Physics, UNSW Sydney, Sydney, New South Wales 2052, Australia

[4]Division of Micro and Nanosystems, KTH Royal Institute of Technology, Stockholm, Sweden

[5]School of Science, Western Sydney University, Locked Bag 1797, Penrith, NSW 2751, Australia

[6]Australian Centre for Nano Medicine and ARC Centre of Excellence in Convergent Bio-Nano Science and Technology, UNSW Sydney, Sydney, New South Wales 2052, Australia

*Corresponding Authors:
A. P. Micolich
Email: adam.micolich@nanoelectronics.physics.unsw.edu.au
D. Mawad
Email: damia.mawad@unsw.edu.au

## Abstract

The organic electrochemical transistor (OECT) with a conjugated polymer as the active material is the elementary unit of organic bioelectronic devices. Increased functionalities, such as low power consumption, can be achieved by building complementary circuits featuring two or more OECTs. Complementary circuits commonly combine both *p*- and *n*-type transistors to reduce power draw. While *p*-type OECTs are readily available, *n*-type OECTs are less common mainly due to poor stability of the *n*-type active channel material in aqueous electrolyte. Additionally, an OECT based complementary circuit requires well matched transport properties in the *p*- and *n*-type materials. Here, a complementary circuit is made using a pair of OECTs having polyaniline (PANI) as the channel material in both transistors. PANI is chosen due to its unique behaviour exhibiting a peak in current versus gate voltage when used as an active channel in an OECT. The PANI based circuit is shown to have excellent performance with gain of ~ 7 and could be transferred on a flexible biocompatible chitosan substrate with demonstrated operation in aqueous electrolyte. This study extends the capabilities of conjugated polymer based OECTs.

**Introduction**

Organic and flexible bioelectronic devices are envisioned to impact future medical applications [1] in clinical diagnostics and therapeutics [2–4]. The organic electrochemical transistor (OECT) is the elementary unit of these devices, with advanced functionality including high signal-to-noise ratio recording or signal amplification [5,6]. Further applications are achieved by building integrated circuits that feature more than one OECT. The simplest example is a complementary logic circuit, which normally features *p*-type (hole majority) and *n*-type (electron majority) transistors in series with their gates connected (Fig. 1A). The operating gate input range is chosen such that when one transistor is 'on' the other is 'off', and vice versa. This makes the output swing between the drive voltage $V_{DD}$ (*p*-transistor on / *n*-transistor off – *p-slope* in Fig. 1A) and ground (*p*-transistor off / *n*-transistor on – *n-slope* in Fig. 1A). The advantage to the complementary architecture is that power draw is reduced by having one of the two devices always in its off state. Complementary circuits are thus widely used in electronics applications with demanding energy supply/dissipation requirements. Both requirements are relevant to bioelectronics, where circuits can require long operation on limited battery supply, e.g., implants, and minimal heat dissipation to prevent tissue/cell damage and device degradation [7]. The complementary circuit described above uses a pair of unipolar devices [8–10]. A more exotic alternative involves a pair of ambipolar devices, where both carrier types can be accessed within the available gate voltage $V_G$ range, typically with some span of insulating behaviour between the *p*-like and *n*-like conduction regions [10,11].

Both unipolar and ambipolar designs pose serious challenges for conjugated polymer semiconductors. Conjugated polymers are predominantly *p*-type materials; the *n*-type polymeric semiconductors synthesized to date have relatively lower carrier mobility than *p*-type [12,13], largely due to intrinsic conformational distortion in the polymer backbone [14]. This mobility mismatch has a detrimental effect on circuit performance.

An additional consideration for bioelectronics is the need for the *n*-type polymer to have an operational electrochemical window compatible with aqueous electrolyte [13]. This further limits the availability of *n*-type materials, with only two reported to date [12,13]. Indeed, only one OECT-based complementary circuit using *n*-type and *p*-type conjugated polymers has been reported so far [12].

The ambipolar design also requires well matched charge transport properties in the *p*- and *n*-type regions [11]. This again limits material choice because in many ambipolar materials the electron and hole transport mobilities differ due to charge-disorder effects [10]. Furthermore, current synthetic techniques are complex and purification of the polymer is very challenging with trace metals occasionally detected in the product [15,16]. Thus, a major quest in the field of bioelectronics has been to find either high performing *n*-type or ambipolar OECT materials with well-matched charge transport, ideally with streamlined synthesis and favourable device processing properties, to further develop OECT-based circuits for applications in neural sensing [17], ion-logic and 'iontronics' [18], and biosymbiotic systems, e.g., electronic plants [19]. Here, we demonstrate the design of OECT-based complementary circuits made using a pair of transistors containing the same channel material (Fig. 1B), which we ultimately engineered to operate on a flexible biocompatible chitosan substrate [20–24] in aqueous electrolyte. As the channel material, we use the well-established conjugated polymer polyaniline (PANI), one of the first organic semiconductors tested in transistors [25].

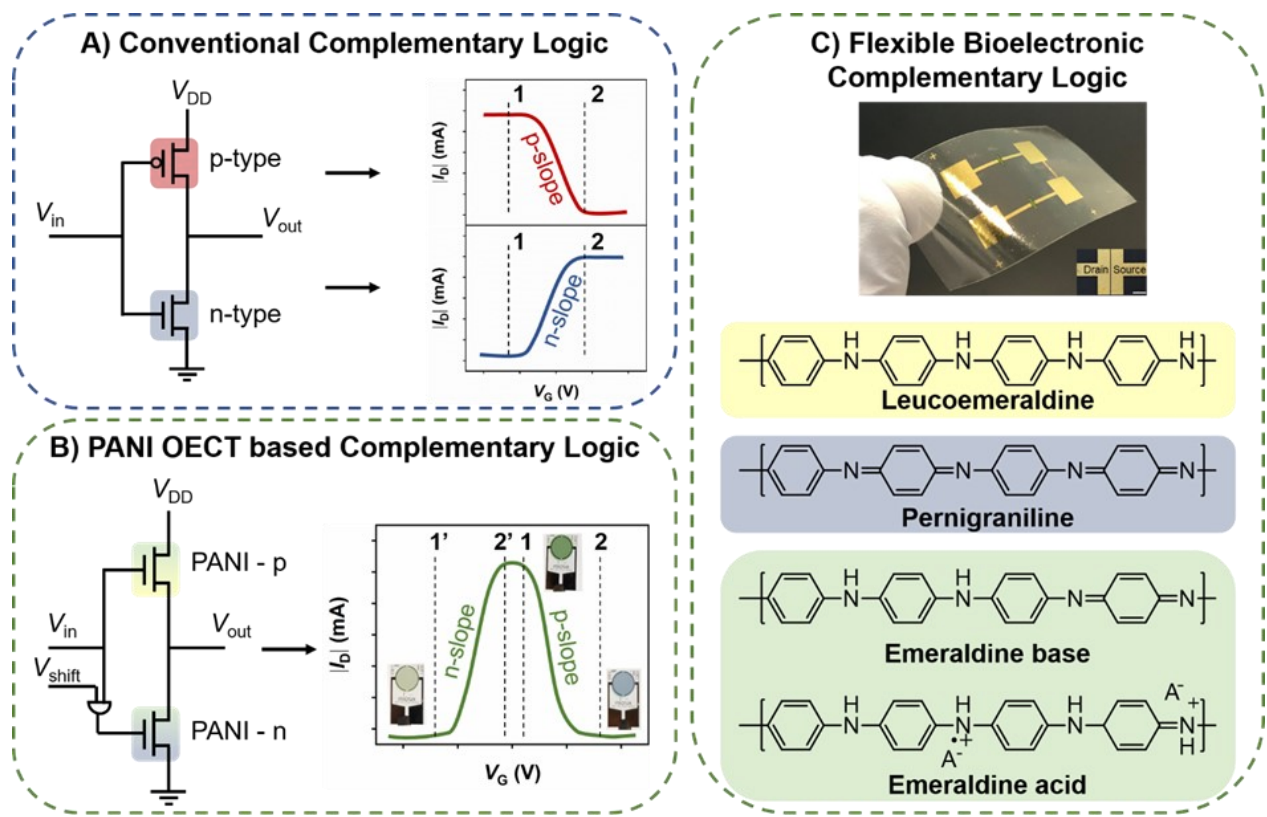

**Fig. 1: Two identical OECT-based flexible complementary circuit. A)** Schematic presentation of a conventional unipolar complementary logic circuit. **B)** Schematic presentation of our PANI OECT based complementary logic circuit. **C)** Digital image of the complementary logic circuit built on a flexible chitosan substrate (top), and chemical structures of PANI in its different oxidation states (bottom). The inset is an optical micrograph of the gold contacts of an individual transistor (scale bar 500 μm).

Our complementary circuit design exploits a known but previously unutilized feature of PANI discovered by Paul *et al.* [25] in 1985. There PANI gave a peak in current versus $V_G$ when used as an OECT channel, in contrast to the monotonic increase/decrease obtained for a conventional unipolar transistor [10]. Circuits designed to cleverly exploit this conductivity peak were not explored at the time; we address this gap here. Our complementary circuit consists of two identical PANI OECTs connected in series between a $V_{DD}$ and ground, as usual. The PANI OECT closest to $V_{DD}$ is operated on the *p*-slope side of the conductance peak (between 1 and 2 in Fig. 1B) by directly applying the input voltage $V_{in}$ to its gate electrode. The PANI OECT closest to ground is operated on the *n*-slope side (between 1' and 2' in Fig. 1B) by using a DC adder to shift the input to its gate electrode by a fixed offset $V_{shift} = -0.4$ V. This arrangement is chosen such that one PANI transistor is 'on' while the other is 'off' enabling complementary circuit performance. We determined the value of $V_{shift}$ from the transfer characteristics as discussed later. A crucial point to note is that polyaniline is not a traditional ambipolar material in this sense – conduction is via hole majority transport across the entire operating range [26,27], and the *p*-like and *n*-like slopes are on opposite sides of a conductivity peak rather than a conductivity valley (i.e., polyaniline = *n*-peak-*p*, traditional ambipolar = *p*-valley-*n*). The peak's symmetry means that the two OECTs in our circuit, operated on opposite sides of the peak, have intrinsically matched characteristics that afford strong potential for complementary circuit use. Using a pair of identical PANI OECTs eliminates the need to develop a second material with matching performance for the small cost of adding a voltage shift apparatus into our input circuit, thereby overcoming a major roadblock to making OECT-based circuits. We also demonstrate that the technology can be transferred onto a flexible biocompatible chitosan substrate operating in an aqueous electrolyte (Fig. 1C), thus paving the way towards the realization of complementary logic circuits based on two identical OECTs for potential applications in organic bioelectronics.

**Results**

**Characterization of our PANI-based OECTs**

For initial studies, PANI OECTs were fabricated on interdigitated gold microelectrodes (IDME) patterned on a glass substrate (PANI-on-IDME OECT). The IDME's serpentine structure gives a channel length of 10 μm and channel width of 50 cm. The PANI film was produced using an oxidative polymerization process with phytic acid as the dopant [28]. The successful deposition of a homogeneous polymer film (thickness 150 nm) was confirmed by atomic force microscopy (fig. S1). The transfer (drain current $I_D$ vs $V_G$) and output ($I_D$ vs source-drain bias $V_{DS}$) characteristics presented in Figs. 2A and 2B were obtained using the ionic liquid 1-butyl-3-methylimidazolium tetrafluoroborate as the electrolyte and a non-aqueous Ag/Ag$^+$ as the gate electrode.

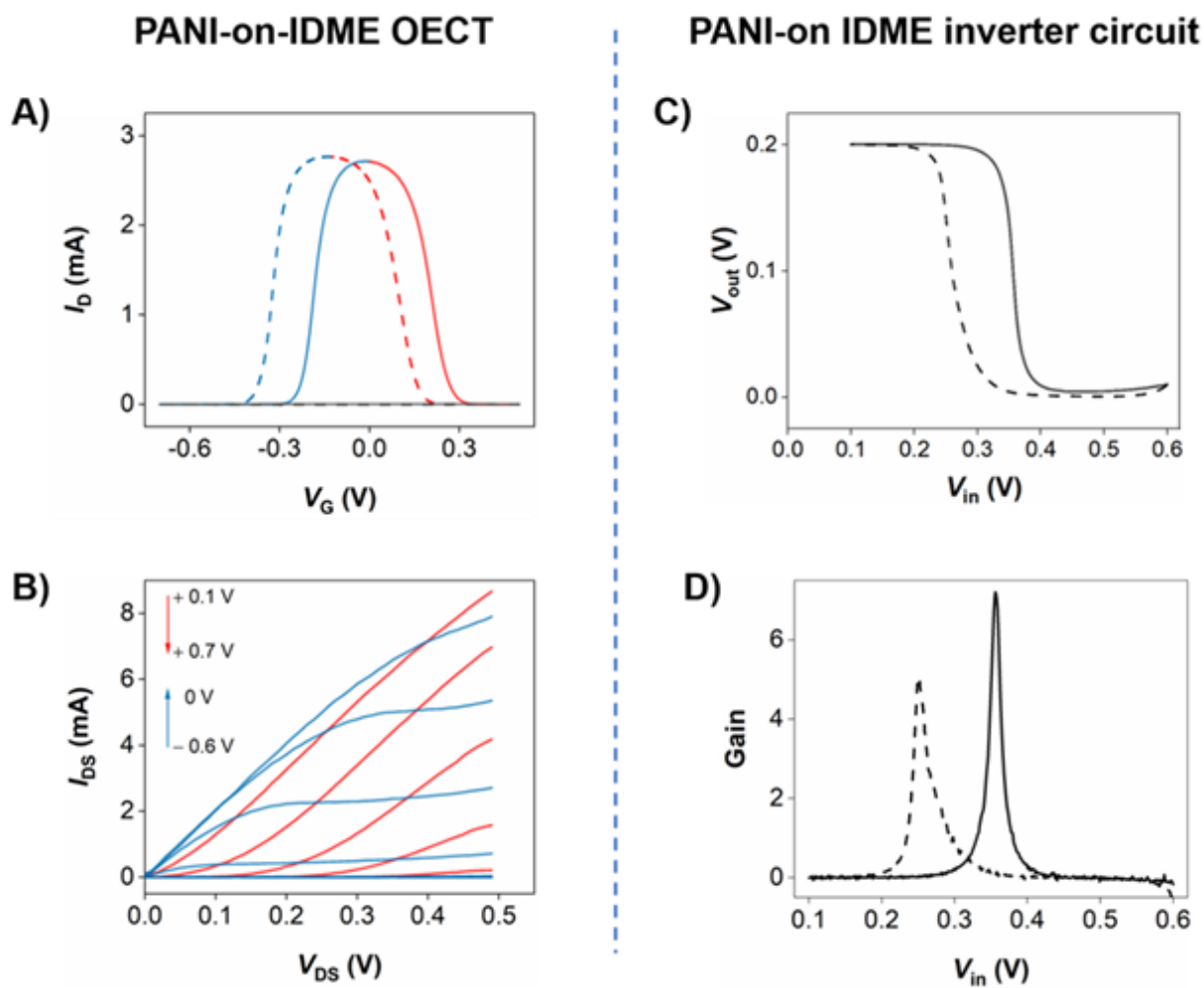

**Fig. 2: PANI-on-IDME OECT as building block for complementary circuit.** A) $I_D$ vs $V_G$ at $V_{DS} = +0.1$ V. Black lines represent the leakage current. B) $I_D$ vs $V_{DS}$ for $V_G$ from −0.6 V to 0 V (blue) and +0.1 V to +0.7 V (red). Data on the *n*-slope and *p*-slope sides of the conductivity peak are presented in red/blue for both A) and B). C) Voltage-transfer characteristic (VTC) for our inverter circuit made using a pair of PANI-on-IDME OECTs connected by external wiring and operating with ionic liquid electrolyte. Green lines represent the VTC adjusted by an offset of −0.19 V. D) Corresponding voltage gain of the inverter. Positive and negative sweeps are presented by solid and dashed lines respectively.

The transfer characteristic sweeping from negative to positive $V_G$ in Fig. 2A (solid line) shows a clear, symmetric peak in $I_D$ versus $V_G$, with a maximum current of 2.43 ± 0.25 mA at $V_G = 0$ V and $V_{DS} = +0.1$ V, corresponding to a channel conductivity of 32.3 ± 0.3 mS cm$^{-1}$. We divide this peak into two regions. The first region (blue) extends from –0.8 V to 0 V, and is where *n*-like behavior is observed, i.e., more positive $V_G$ increases the conductivity. The second region (red) extends from 0 V to +0.5 V, and is where *p*-like behavior is observed, i.e., more positive $V_G$ decreases the conductivity. We observe the same two regions as we sweep towards more negative $V_G$, however with the peak current shifted by −100 mV. This hysteresis correlates with the hysteresis observed in the cyclic voltammetry (fig. S2) and is in agreement with the literature [29–31]. The hysteresis has been described for conjugated polymer-based devices, in which the polymer has been highly oxidized without causing its degradation [29,30]. The hysteresis is attributed to the transition between different electronic states and the charge density of the oxidized polymer. Of note, recent designs have introduced hysteresis in conjugated polymer OECT devices to enable their application in memory devices [32–34]. The output characteristics in Fig. 2B show behavior consistent with the transfer characteristics in Fig. 2A. We repeated these measurements for our PANI-on-IDME OECT at $V_{DS} = -0.1$ V (fig. S3A and S3B). Similarly, a current peak is observed as $V_G$ is increased towards positive potential.

**Characterization of PANI-based OECT complementary logic circuits**

We set out to investigate whether a single-material complementary circuit can be built using two identical OECTs, i.e., featuring the same active channel material PANI, based on the *n*-like and *p*-like behavior we have obtained from our PANI-based OECT. Figure 2C shows the voltage-transfer characteristic (VTC) for our inverter circuit (Fig. 1B) with $V_{DD} = +0.2$ V and offset voltage $V_{shift} = -0.4$ V applied to the OECT closest to ground. We determined the value of $V_{shift}$ from the transfer curves of the single OECTs used in the circuit, such that we obtain one OECT in its 'on' state and the other in its 'off' state (fig. S4). There is a slight offset between the sweep from negative $V_{in}$ to positive $V_{in}$ (solid line) and sweep back from positive $V_{in}$ to negative $V_{in}$ (dashed line), consistent with the hysteresis previously shown in Fig. 2A. Fig. 2D shows the DC gain $\partial V_{out}/\partial V_{in}$ versus $V_{in}$ with a peak gain of 7.2 obtained at the switching threshold $V_M = +0.36$ V. Our inverter gain is > 1, meeting an essential functional criterion of an inverter [10]. It also has higher gain compared to the only conjugated polymer OECT based complementary circuit reported to date (~ 4) and measured at the same $V_{DD}$ [12]. The ideal switching threshold for well optimized complementary inverter circuits is $V_{DD}/2$, and deviation of our inverter from this could be for two key reasons. First, the switching thresholds are not symmetric about 0 V, i.e., $V_{Th,n} \neq -V_{Th,p}$, where $V_{Th,n}$ and $V_{Th,p}$ are the *n*-type and *p*-type switching thresholds (fig. S4C), respectively. Second, the quantity $k = \mu C(Z/L)$, where $\mu$ is mobility, $C$ is gate capacitance, $Z$ is channel width and $L$ is channel length, is not perfectly balanced for the two transistors [10], as indicated by the different magnitude in saturation currents exhibited by each device (fig. S4C). Our PANI OECT-based complementary logic circuit can also function as an amplifier circuit when operated at a negative $V_{DD} = -0.2$ V as shown in Figures S2C and S2D. Although there is some room for improved performance through optimization, our work here represents the first proof-of-concept for an inverter circuit built from two OECTs featuring identical active channel material.

**Simultaneous transistor, electrochemical and optical characterization of our OECTs**

PANI is a *p*-doped semiconductor in the presence of phytic acid. The origin of the transition between the *n*-like and *p*-like behavior we observe here is electrochemically driven. We show this by investigating the electrochemical and optical properties of PANI using an in-house built apparatus **(Fig. 3A and fig. S5)** that enables simultaneous transistor, CV and UV-Vis measurements of a PANI-on-IDME as used in our OECT. The set-up is described in the supplementary information. The intent here is not to replace properly engineered CV and UV-Vis instruments but to augment them – we use the data obtained from this set-up to ensure that measurements taken using our CV and UV-Vis systems are correctly aligned in gate potential with the transistor characteristics.

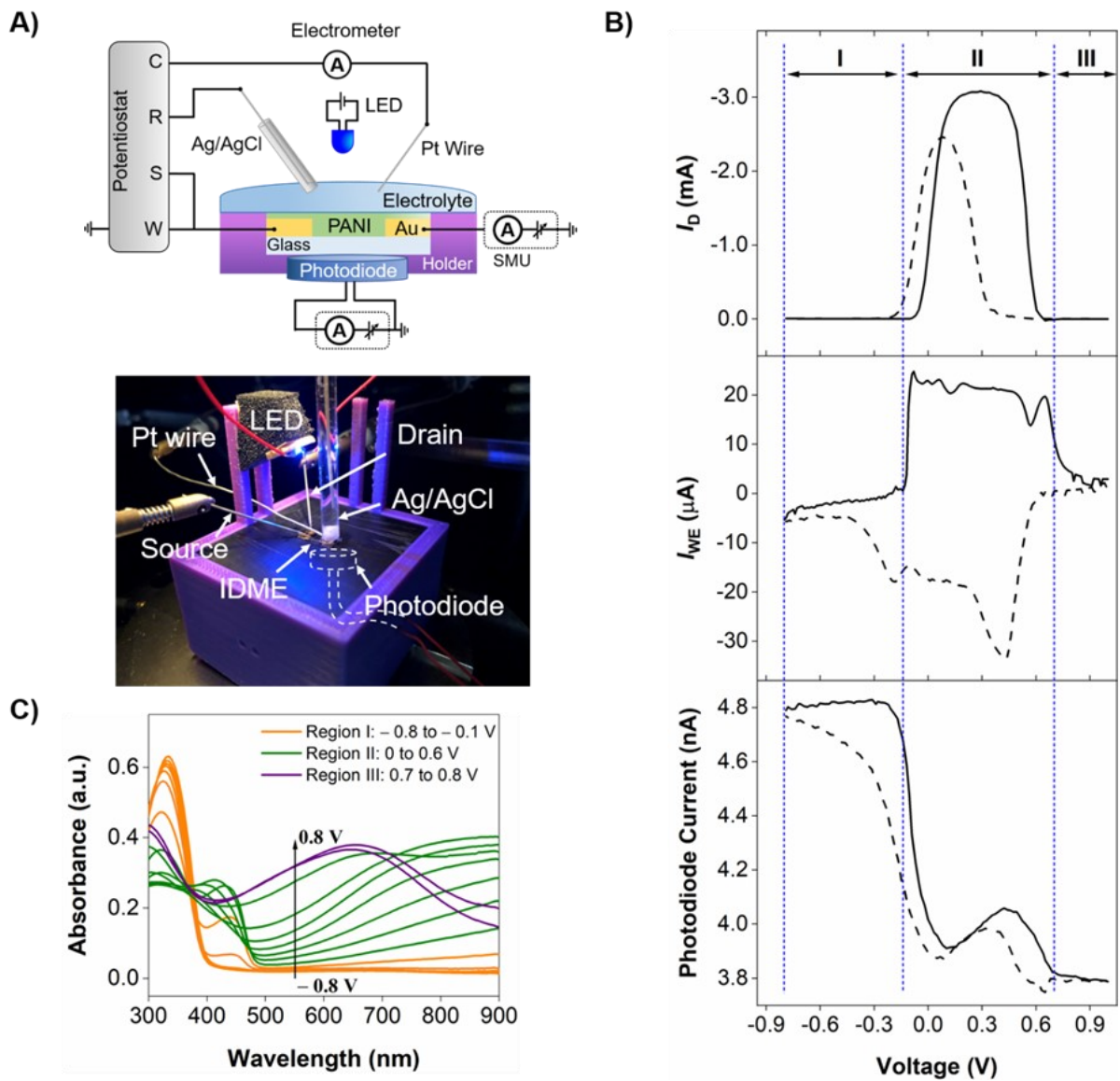

**Fig. 3**: *In situ* **monitoring of the redox changes in PANI OECT during operation.** A) Schematic (top) and photograph (bottom) of the apparatus used for near-simultaneous measurements of transfer characteristics, cyclic voltammetry and UV-visible absorption. B) Plots of (top) $I_D$ at $V_{DS}$ = –0.1 V, (middle) $I_{WE}$ working electrode (WE) current and (bottom) photodiode current under 428 nm illumination versus $V_G$ swept at 50 mVs$^{-1}$. C) *In situ* spectroelectrochemistry of PANI film on ITO-coated glass as a function of applied potential between –0.8 and +0.8 V.

Figure 3B presents measurements of $I_D$ (top), working electrode current $I_{WE}$ (middle) and photodiode current (bottom) versus $V_G$. The PANI clearly exhibits different characteristics defined by three distinct regions of potential: Region I (–0.8 to –0.1 V), Region II (–0.1 to +0.7 V), and Region III (+0.7 to +1.0 V). The cyclic voltammogram (middle) shows typical redox process in which PANI changes oxidation states between leucoemeraldine (non-conductive and fully reduced – Region I), emeraldine (conductive – Region II) and pernigraniline (non-conductive and fully oxidised – Region III) [35,36]. In Regions I and III, the Faradaic current is negligible, corresponding to the off states of the PANI OECT in the transfer characteristic. In contrast, the Faradaic current is significantly higher in Region II, indicating that the PANI is in the conductive form. The on state of the OECT coincides directly with Region II.

We concurrently measured the optical transmission via the photodiode current (Fig. 3B bottom) since absorption spectroscopy enables the oxidation states of PANI (Fig. 1C) to be identified [36]. Reduced photodiode current in Fig. 3B corresponding to enhanced absorption. When PANI is in its leucoemeraldine state, only the 320 nm peak is observed. The emeraldine acid shows a peak at 430 nm accompanied by a broad shoulder centred at > 800 nm. A blue shift of the polaron region to ~ 650 nm or lower identifies the emeraldine base and pernigraniline states. Since the emeraldine acid is the most electrically conductive form of PANI, we chose to focus on the absorption at 428 nm (Fig. 3B) and 850 nm (fig. S6). In Fig. 3B, the photodiode current remains constant and at maximal value for $V_G < -0.22$ V. We observed a sharp decrease in photodiode current at the boundary between Regions I and II. We attribute this sharp increase in absorption to protonation of the leucoemeraldine state to form the emeraldine acid state. We confirm this process with separate spectroelectrochemistry measurements shown in Fig. 3C. As we move from Region I to Region II, we observe the weakening of the 320 nm peak corresponding to the leucoemeraldine state and strengthening of the 430 nm peak corresponding to the emeraldine acid state, indicating a transition between the two oxidation states. We observe fluctuations in the photodiode current with further increase in $V_G$ (Regions II and III; Fig. 3B), but it remains low compared to its initial value at $V_G < -0.22$ V. The fluctuations indicate further chemical change in the film; the emeraldine acid co-exists with the emeraldine base in Region II and converts to the pernigraniline state in Region III. The blue shift in the polaron region and decrease in 430 nm peak in Fig. 3C further support this conclusion. Combining optical transmission with cyclic voltammetry while the OECT was operating enabled us to identify with certainty that the non-conductive leucoemeraldine and pernigraniline states are the dominant species in Regions I and III respectively, where the OECT was in the off state, whilst the emeraldine state was dominant in Region II, where the OECT is in the on state.

**Density functional theory study of conduction states in our PANI OECTs**

We performed density functional theory (DFT) calculations for PANI in different oxidation states to further understand Regions I-III. In previous studies [37,38], analysis of the electronic states was performed for isolated PANI polymers, i.e., PANI surrounded by vacuum. Here, we explicitly simulate and analyze PANI in its different oxidation states while in contact with the Au electrode, an integral component in the OECT device. This is an original aspect of our simulations in comparison to the literature. To perform this type of simulation properly, we have carried out very intensive DFT+D calculations for a large system composed of 128 Au atoms, 52 organic ions and a large vacuum region of 30 Å of length. We reveal, that in the absence of any defects, PANI was physisorbed on the gold substrate via hydrogen bonds of the Au⋯H-X type (fig. S7) [39]. The main interactions between the PANI polymer and the Au substrate are long-range dispersion-like with no considerable effect on the electronic band structure of the polymer.

We find that the leucoemeraldine base is strongly insulating owing to a wide band gap of 2.15 eV and a Fermi energy just above the valence band edge (fig. S7A) in agreement with the experiments. For the emeraldine acid state, we find empty electronic states at ~ 0.1 eV above the valence band edge. These arise from polaron formation upon protonation of the leucoemeraldine polymer backbone. These empty states are sufficiently close to the valence band edge so that charge carriers can be excited upon application of a small electric bias making PANI an effective conductor (fig. S7B). For the pernigraniline base, we find that the density of polaron states just above the valence band edge is increased as compared to the emeraldine state (fig. S7C). Upon application of an appropriate electric field, it is possible to fill up such unoccupied electronic states and attain an insulator state characterized by an energy band gap of 1.55 eV. Overall, these calculations compare well with others in the literature [37,38,40], and support the argument that the redox changes occurring in PANI lead to the conductance peak in the transfer characteristics.

**Monolithic PANI-OECT complementary circuits on flexible biocompatible substrates operating in aqueous electrolyte**

Low operating voltage is a vital property for building bioelectronic circuits, which inevitably need to operate in an aqueous electrolyte environment. An attractive aspect of PANI is that its electrochemical activity is within an ideal range for this purpose (< 1 V) (fig. S2). We have also previously shown that phytic-acid-doped PANI can be polymerized on a flexible biocompatible chitosan substrate to form an electronically and mechanically stable conductive patch, which we developed for electrically bridging infarcted cardiac tissue [28]. That put us in an ideal position to directly transfer our PANI OECT complementary circuits from non-monolithic glass IDMEs, i.e., separate glass IDME chips wired in an external circuit, to fully monolithic integrated on flexible biocompatible chitosan substrates [41–43] and actuated by aqueous electrolyte (Fig. 4A). We show the transfer characteristics in Fig. 4B for a single PANI-on-chitosan OECT obtained using 0.1 M NaCl aqueous electrolyte with an Ag/AgCl electrode as the reference electrode for supplying $V_G$. The conductance peak is shifted to slightly lower $V_G$ (−0.08 V) compared to our devices on glass (Fig. 2A), which may simply arise from differences in substrate charge between glass and chitosan. The peak current (0.63 mA) is lower for the same $V_{DS}$, which is more related to the change in transistor geometry than an intrinsic change in PANI properties.

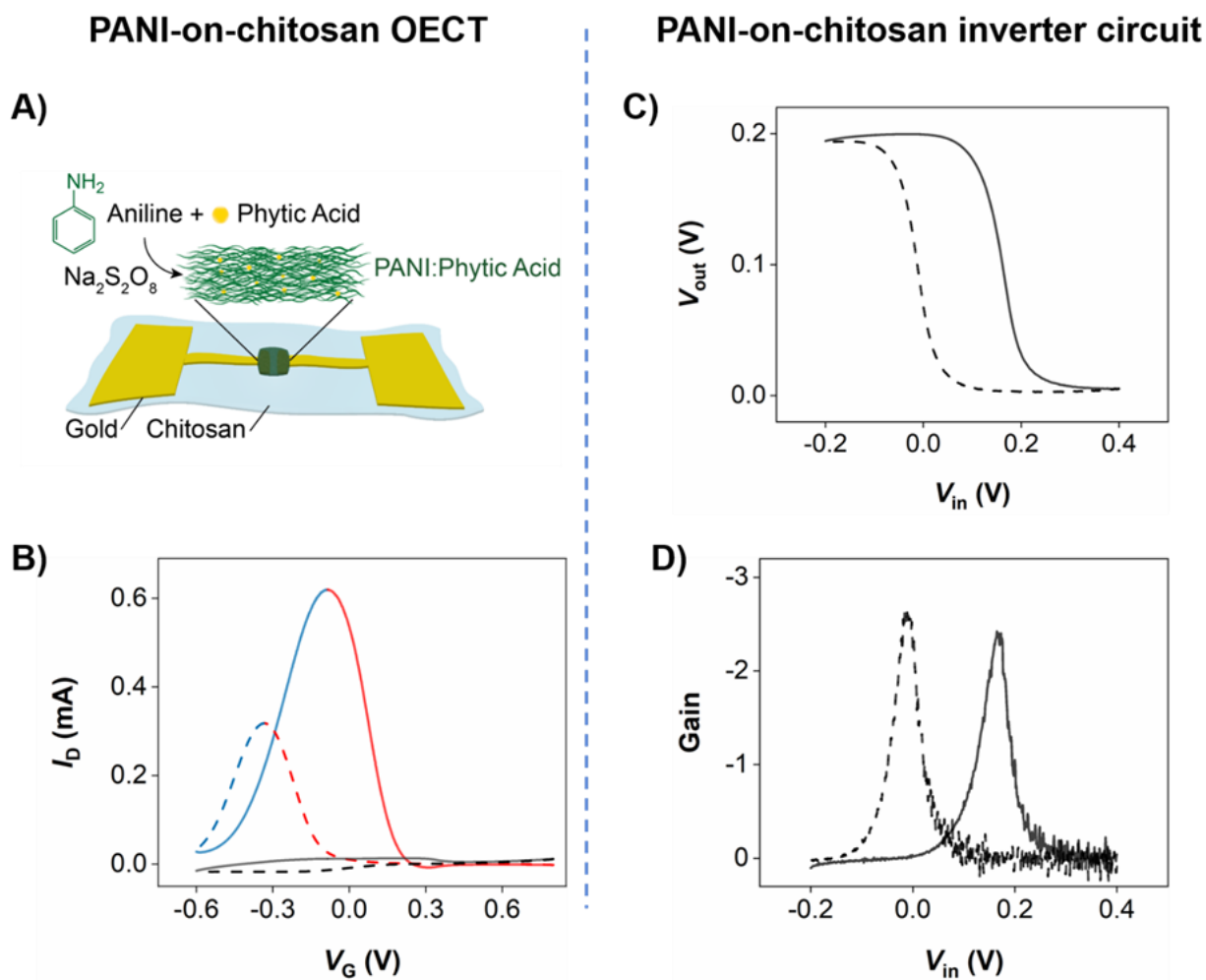

**Fig. 4: A flexible PANI-on-chitosan OECTs complementary circuit.** A) Schematic of the oxidative polymerization of PANI doped with phytic acid on a chitosan film with patterned Au electrodes on the surface. B) $I_D$ vs $V_G$ at $V_{DS}$ = +0.1 V. Black lines represent the leakage current. C) VTC for inverter circuit made on flexible chitosan film operating with aqueous NaCl electrolyte. D) Corresponding voltage gain of the inverter. Positive and negative sweeps are presented by solid and dashed lines respectively.

Fig. 4C and 4D show VTC and corresponding gain for our PANI-on-chitosan inverter circuit (Fig. 1C) with $V_{DD}$ = +0.2 V and offset voltage $V_{shift}$ = –0.4 V applied to the OECT closest to ground. We observe hysteresis similar to our PANI-on-IDME inverter and in agreement with the hysteresis recorded for the single PANI-on-chitosan OECT (Fig. 4B). We obtain a peak gain of 2.4, lower than for our PANI-on-IDME circuit, which could arise either from a less optimal transistor geometry or trapping of electrolyte ions in the positively charged chitosan. Transfer, output and VTC characteristics of the PANI-on-chitosan circuit operating at a negative $V_{DD} = -0.2$ V are shown in fig. S9.

**Discussion**

Introducing electronic circuits into bioelectronic devices will enable new and improved biomedical functionalities. Conjugated polymer based OECTs have demonstrated superior performance at the biointerface, in comparison to other transistor architectures such as the organic field-effect transistor (OFET) or the electrolyte-gated OFET [44]. This is due to the mixed electronic and ionic transport that OECT displays. Despite the fact that transistors are rarely used alone and often built into integrated circuits for higher function, OECTs have been largely explored as single devices. Few attempts have been recently made towards modifying or using the OECT as an elementary unit in building advanced electronic circuits [44–47], with only one example reporting a conjugated polymer based complementary circuit [12]. We speculate that the lack of advancement in developing complementary circuits based on conjugated polymers is due to the limited choice of materials. Conjugated polymers are mostly *p*-type with very few *n*-type materials readily available [14]. Here, we overcome this limitation by taking a radical new approach towards building a complementary circuit. We demonstrate that a single material complementary circuit can be built from two identical PANI-based OECTs, eliminating the need for an *n*-type conjugated polymer. We achieve this by arranging the two identical OECTs in an inverter circuit and applying a small additional voltage to offset the operating point of one OECT relative to the other, thereby giving dc gain as high as 7.2. We transfer the technology on a chitosan substrate to build a flexible inverter with demonstrated functionality in aqueous electrolytes. As a proof-of-concept with unique characteristics, further work is required to optimize the performance of our device. Dedicated studies maybe performed to probe whether the observed hysteresis can be minimized by controlling the charge density on the oxidized polymer (degree of doping). Our PANI-based OECT was fabricated by chemically polymerizing aniline on the electrode surface, accounting for slight variations in device geometry. The fabrication process can be improved by using photolithography or electropolymerization that will enable control over the film thickness [48]. Minimizing variations in the geometry of devices will achieve equivalent saturation currents and shift the inverter response to ideal behavior.

Our work's key novelty is the application of the conductance peak in PANI OECT transfer characteristics to realize OECT-based complementary circuits using only a single common channel material. This was made possible because of the finite electrochemical window of PANI that is within a potential range suited for operation in water (< 0.9 V). The conductance peak we observe in our PANI OECTs is consistent with literature [25,27,30,49], and similar behavior has been shown for other conjugated polymers such as polypyrrole and polythiophenes, though for potentials > 1 V [29,50,51]. However, this is the first time this feature has been exploited for novel complementary circuit design featuring two OECTs with identical channel composition. Furthermore, our design approach to integrate the inverter circuit in a flexible biocompatible substrate enabled its operation in aqueous electrolyte. We have previously shown that a conductive patch made from PANI-on-chitosan electrocouples with damaged cardiac tissue increasing conduction velocity (~ 25%), and is not pro-arrythmogenic following *in vivo* implantation [28]. Our work set the foundation for the flexible complementary logic circuit we here developed for *in vivo* operation at the biointerface. The next step will be to test the capability of the circuit to process signals of physiological frequencies, towards its use as a real-time processing unit at the biointerface [52]. The flexible inverter can also have applications in memory and neuromorphic devices [6,32,53], exploiting the hysteresis shown in our PANI OECT devices. The present work sheds a new light on conjugated polymers, extending their capabilities beyond use in a single OECT device but towards single material channels in complementary logic circuits.

**Experimental Methods**

*Materials.*

All chemicals were purchased from Sigma Aldrich and used without any further purification. Any water used was deionized to 18 MΩ.cm using a Millipore (Milli-DI) pro system unless otherwise specified.

*PANI-on-IDME.*

OECTs were fabricated on interdigitated microelectrode (IDME) structures purchased from Micrux Technologies (ED-IDE1-Au). PANI films were obtained by oxidative polymerization of aniline directly on the IDME as described previously [28,54]. Briefly, Solution 1 was prepared by mixing 470 μL phytic acid and 230 μL aniline in 1 mL of water. Solution 2 was prepared by dissolving 143 mg ammonium persulfate in 0.5 mL of water. Then, 7.7 μL of Solution 1 and 2.3 μL of Solution 2 were mixed in an Eppendorf tube and immediately transferred onto the IDME. The IDME was left for 3 hours to allow for PANI polymerization. The reaction was quenched by rinsing with water until no PANI flaked off the surface.

*PANI-on-chitosan OECT.*

Chitosan substrates (20 μm thick) were prepared by dropcasting 1.6 mL of chitosan solution (1.5wt% chitosan in 2wt% acetic acid aqueous solution) on a microscope slide (76 mm x 25 mm).[55] Electrodes were deposited using a PVD75 vacuum thermal evaporator system with a shadow mask. The metallization was 5 nm Ti and 100 nm Au, with the Ti used to ensure good metal adhesion to the chitosan. The transistor channel had width 20 μm and length 2 mm. PANI was chemically polymerised in the conductive channel using the method previously described. This resulted in strong adhesion of PANI on the substrate due to the strong chelation between the negatively charged phytic acid and the positively charged PANI and chitosan.

*Atomic force microscopy.*

IDME topography before and after PANI deposition was acquired using a JPK Instruments atomic force microscope with NanoWizard II software. The microscope was operated in contact mode in air using silicon cantilevers (AppNano ACT-50; Length 125 μm, width 30 μm, tip radius <10 nm) with 36 Nm-1 nominal spring constant and 200-400 kHz resonant frequency. Scanning rate was 1 Hz with pixel resolution 1024 x 1024.

*In-situ UV-Vis spectroelectrochemistry.*

PANI was polymerized on ITO slides (60 mm x 8 mm) to serve as the working electrode and put into a 3 mL cuvette containing 1-butyl-3-methylimidazolium tetrafluoroborate ionic liquid. A platinum wire and Ag/AgCl electrode in 0.1 M AgNO3 (CHI Instruments) were immersed in the ionic liquid as counter and reference electrodes, with potential applied and current measured using an Ivium Vertex 100 mA potentiostat. The absorption spectra were obtained with a PerkinElmer Lambda 900 UV-Vis spectrometer over the 300 nm to 900 nm wavelength with 5 nm steps.

*Electrical characterization of the OECT.*

Transistor characteristics were measured using a pair of Keithley 2401 Source Measure Units (SMUs) and a Keithley 6517A electrometer. The first SMU supplies the drain-source voltage VDS and measures source current IS, and the second SMU supplies the gate voltage VG and measures gate leakage current $I_G$ (via Ag/AgCl in 0.1 M $AgNO_3$ electrode; CHI Instruments). The electrometer measures the drain current ID flowing to virtual ground. Voltage sweeps were performed with a step of 1 mV. The electrical characteristics of the device on IDME and chitosan substrate were obtained using 1-butyl-3-methylimidazolium tetrafluoroborate ionic liquid and 0.1 M NaCl solution respectively. Conductivity of the channel (σ) was calculated using the following formula:

$$\sigma = w/(R \cdot L \cdot t)$$

where R is the resistance, L and w are the length and width of the channel respectively, and t is the thickness of the PANI film.

*Electrical characterization of the OECT-based complementary circuit.*

The inverter supply voltage $V_{DD}$, DC input voltage $V_{in}$ and shift voltage $V_{shift}$ were applied using three Keithley 2401 source-measure units. The output voltage $V_{out}$ was acquired using a Keithley 2000 Multimeter. For AC measurements, a bias of VDD was applied using a Keithley 2401 source measure unit. The AC input voltage $V_{in}$ was applied using a signal generator DS345 (Stanford Research System) and measured via the USB-6216 Acquisition Device. The output signal Vout was measured and recorded using a preamplifier Femto DLPVA and a National Instruments USB-6216 Data Acquisition Device. An Ag/AgCl in 0.1 M $AgNO_3$ electrode immersed in 1-butyl-3-methylimidazolium tetrafluoroborate was used for the logic circuit with PANI-on-IDME transistors and an Ag/AgCl in 0.1 M $AgNO_3$ electrode immersed in 0.1 M NaCl solution was used for the chitosan-based logic circuit.

*Concurrent measurements of optical and electrochemical properties.*

CV and UV-visible absorption spectroscopy (UV-Vis) are generally used to investigate the electrochemical and optical properties of PANI films. These are commonly performed on separate films/substrates under different experimental conditions using different instruments, making comparative analysis more challenging. To address this, we built an apparatus (shown in Fig. 3A) that enables transistor, CV and UV-Vis measurements to be obtained from our PANI-on-IDME in exactly the form they are used in our OECT under identical conditions and nearly simultaneously. The apparatus consists of a box with a transparent upper window containing a 15 $mm^2$ Si photodiode (Centronic OSD15-5T) mounted directly underneath the 3.5 mm diameter active region of the glass IDME chip and used to measure transmitted light, which is proportional to induced photocurrent. The IDME was illuminated obliquely from above by an LED with wavelength 428 nm or 850 nm to accommodate the Ag/AgCl microelectrode contacting the ionic liquid electrolyte. The transistor and CV measurements were obtained with a hybrid set-up consisting of an Ivium Vertex potentiostat, a Keithley 6517A electrometer and a Keithley 2401 source-measure unit. For conventional CV measurements, the source and drain would be both connected to the working electrode, with the reference (Ag/AgCl) and counter (Pt) electrodes immersed in the ionic liquid sitting on the PANI transistor channel. Our set-up in Fig. 3A differs in two ways from this. First, only the drain electrode remains connected to the working electrode input for the potentiostat; the source electrode is instead connected to the Keithley 2401 to enable a source-drain bias $V_{SD}$ to be applied. Second, a floating Keithley 6517A is put into the counter electrode line. This enabled us to measure the corresponding gate leakage current $I_G$ flowing via the counter electrode when a gate voltage $V_G$ is applied on the reference electrode. Note that this $V_G$ is identical to the voltage (V) applied in cyclic voltammetry, and the gate leakage current $I_G$ is equivalent to the current measured at the working electrode $I_{WE}$ in a typical three electrode set-up. The cyclic voltammetry presented in Fig. 3A was obtained by plotting $I_{WE}$ versus $V_G$. The potentiostat working electrode measures drain current $I_D$ and the Keithley 2401 enables us to also measure source current $I_S$ whilst sourcing $V_{SD}$, giving us full details for transistor characterization. For a pure CV trace, we would run $V_{SD}$ = 0 V, with $I_G = I_S + I_D$ giving us the current $I$ conventionally measured in this technique. We have confirmed this combination of instruments does not adversely affect outcomes of either characterization, i.e., it gives transistor characteristics matching those we obtained in our normal transistor characterization circuit and likewise for CV compared to the Ivium used alone.

***Density functional theory calculations.***

First-principles calculations based on density functional theory (DFT) [56] were done with the VASP code [57], using the generalized gradient approximation to the exchange-correlation energy developed by Perdew *et al.* [58]. Possible dispersion interactions in the system were captured with the D3 correction scheme developed by Grimme *et al.* [59]. The projector-augmented-wave method was used to represent the ionic cores and the following electronic states were considered as valence: Au 5$d$-6$s$, C 2$s$-2$p$, O 2$s$-2$p$, and H 1$s$. Wave functions were represented in a plane-wave basis truncated at 750 eV and a force tolerance of 0.01 eV Å$^{-1}$ was imposed in the geometry optimizations. The energies of the optimized systems converged to within 2 meV per atom using the parameters above and a k-point grid of 1 x 2 x 2 for integration within the Brillouin zone. Large simulation boxes typically containing 180 atoms were employed to simulate the Au substrate/PANI systems (128 atoms for the gold substrate and 52 for the PANI). Periodic boundary conditions were applied along the three Cartesian directions and a vacuum region of ~ 30 Å was considered in the simulation cell to avoid spurious interactions with the neighboring image systems (fig. S7). We checked that our simulation DFT results were appropriately converged with respect to the adopted PANI length of four carbon aromatic rings by carrying out additional calculations for longer polymers containing six monomer units.

**Acknowledgements**

The work was performed in part using the NSW node of the Australian National Fabrication Facility at UNSW Sydney. E.Z. thanks the Swedish Research Council (VR International Postdoc Grant #2017-06381) and the Royal Swedish Academy of Sciences (Kungl. Vetenskapsakademiens stiftelser, #LN2017-0035) for funding.

**Funding:** This work was funded by the Australian Research Council under DP170104024 and DP190102560.

**Author contributions:** D.M. and A.M. conceived the research with L.T. help. L.T. performed all the experiments. C.C. performed the DFT calculations. A.L. supervised the AFM measurements. L.T., A.M. and D.M. performed the data analysis and data interpretation. All authors contributed to the manuscript writing.

**Competing interests**: The authors declare that they have no competing interests.

**Data and materials availability:** All data needed to evaluate the conclusions in the paper are present in the paper and/or the Supplementary Materials. Data can be obtained upon request from the authors.

**Supplementary Materials for**

**Flexible complementary logic circuit built from two identical organic electrochemical transistors**

Lorenzo Travaglini, Adam P. Micolich, Claudio Cazorla Silva, Erica Zeglio, Antonio Lauto, Damia Mawad

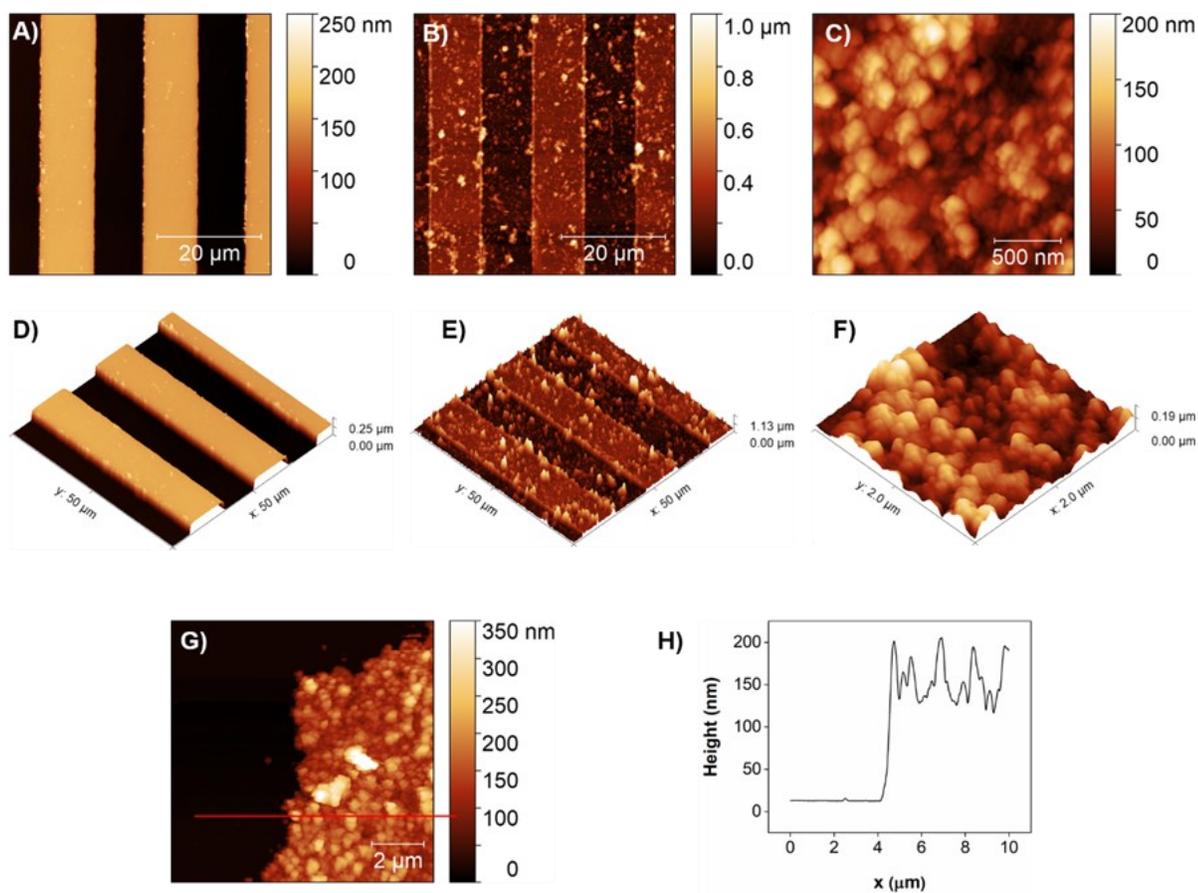

**Fig. S1. AFM surface topography of PANI-on-IDME.** AFM images of bare electrodes **(A)**, and electrodes with an electroactive PANI film on the surface, 50 x 50 μm **(B)** and 2 x 2 μm **(C)**. **D-E-F)** 3D rendered AFM topography of the same areas. AFM image of the edge of PANI film **(G)** and its relative thickness profile **(H)**.

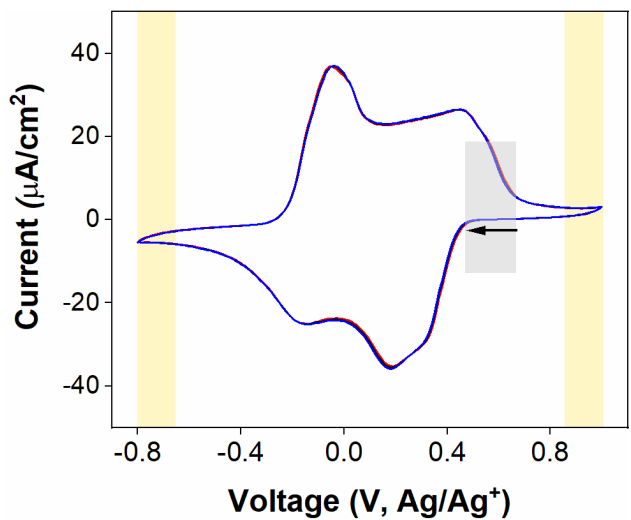

**Fig. S2. Cyclic voltammogram (CV) of PANI on IDME.** Three cycles were recorded in ionic liquid at a scan rate of 50 mV.s$^{-1}$. The counter electrode was a Pt plate and the reference was Ag/Ag$^+$ electrode. The black arrow and grey box highlight hysteresis as we sweep the potential back from positive to negative, in correlation with the hysteresis observed in the transfer curves. The yellow regions indicate the edges of the finite electrochemical window of PANI that does not exceed 1 V.

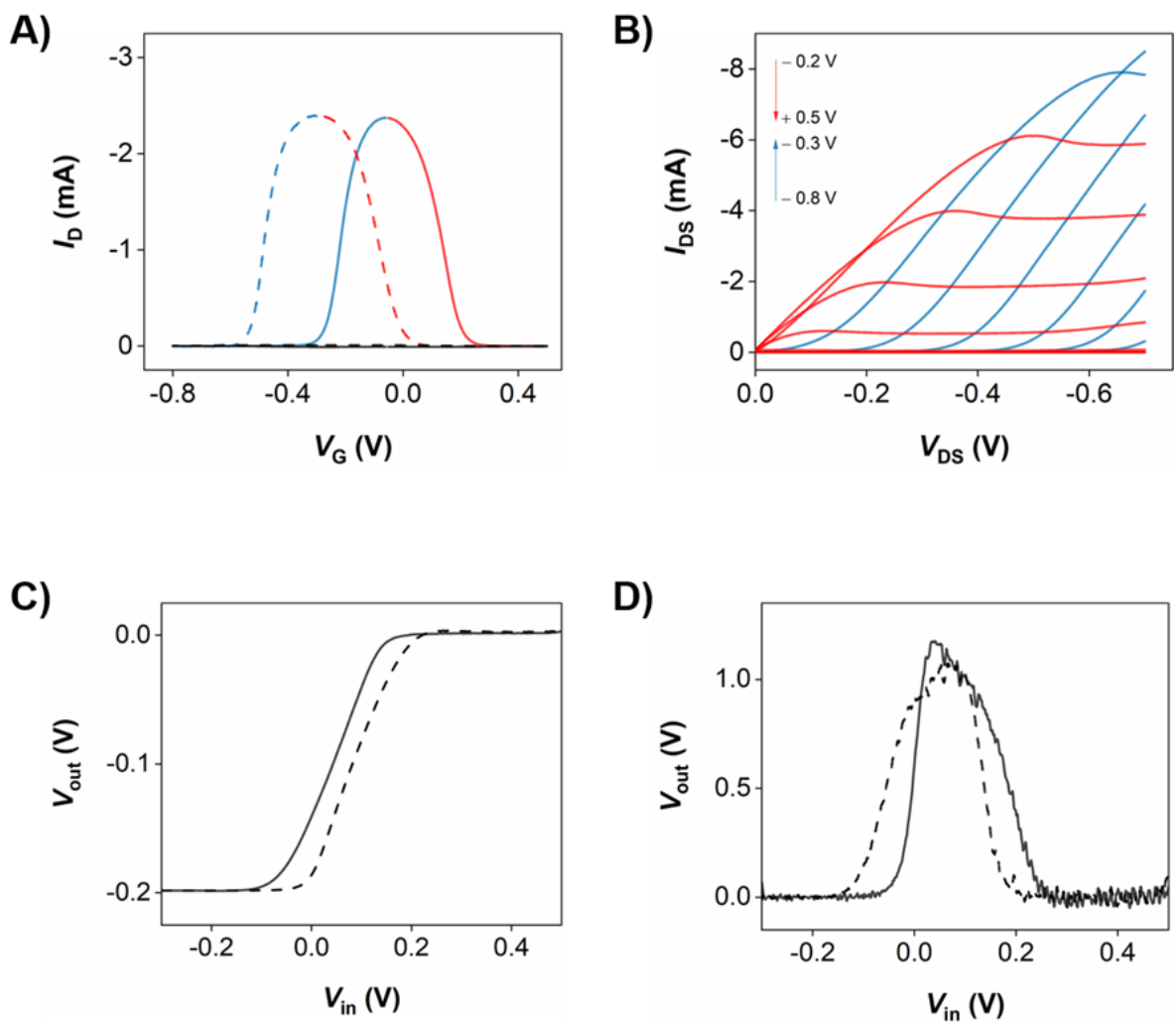

**Fig. S3. Electrical characteristics of PANI-on-IDME OECT. A)** Drain current $I_D$ vs gate voltage $V_G$ at drain-source voltage $V_{DS} = -0.1$ V. Black lines represent the leakage current. **B)** $I_D$ vs $V_{DS}$ for $V_G$ from −0.8 V to −0.3 V (blue) and −0.2 V to +0.5 V (red). Data on the *n*-slope and *p*-slope sides of the conductivity peak are presented in red/blue for both **A)** and **B)**. **C)** Voltage-transfer characteristic (VTC) for amplifier circuit made using a pair of PANI-on-IDME OECTs connected by external wiring and operating with ionic liquid electrolyte. **D)** Corresponding voltage gain of the amplifier.

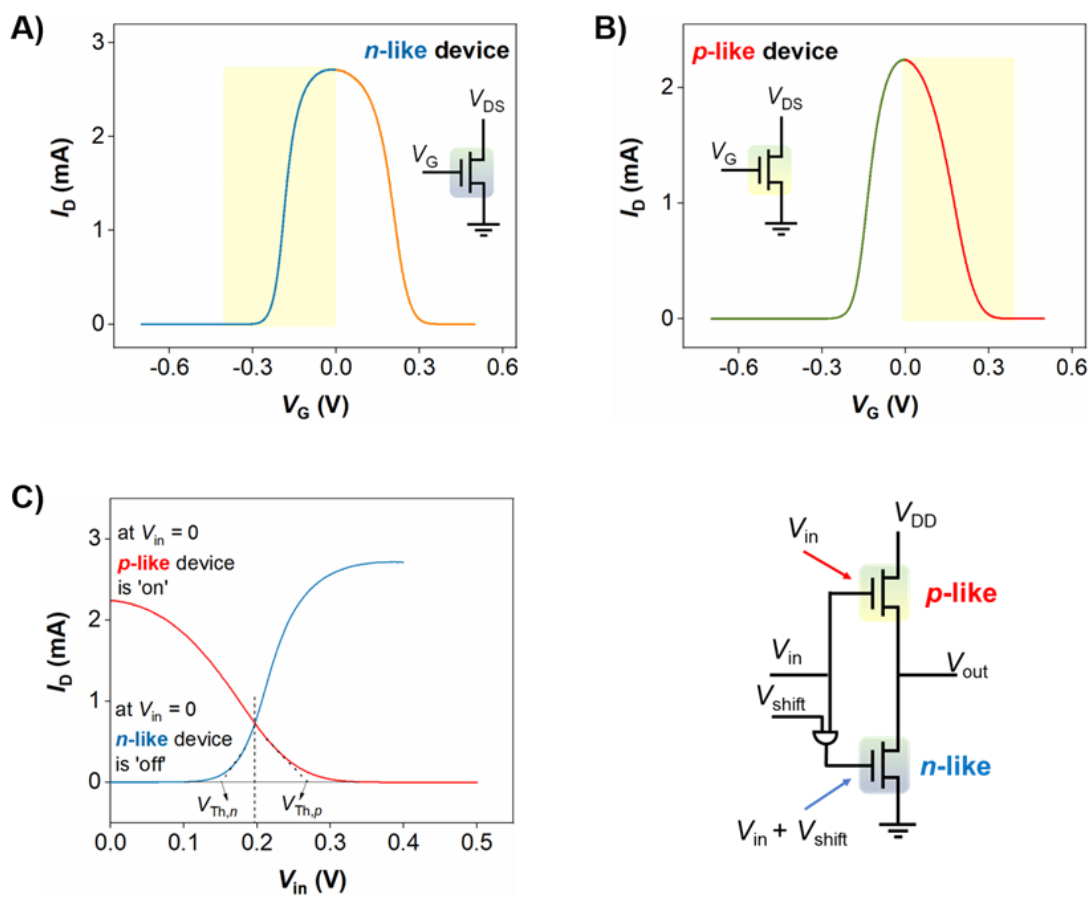

**Fig S4. Applying a $V_{shift}$ to enable complementary circuit performance. A and B)** Transfer curves of the two PANI-based OECTs used to build the inverter circuit. Regions highlighted in yellow are the "active regions" we use in the circuit. The orange and green part of the curves are the "inactive regions". **C)** $I_D$ of each single OECT device plotted versus the input voltage $V_{in}$ applied in the complementary circuit. The threshold voltages, $V_{Th,n} = 0.15$ V and $V_{Th,p} = 0.27$ V, were found to be centred about 0.20 V. The scheme in (C) represents the circuit featuring two identical PANI OECTs but with a $V_{shift}$ applied to the $n$-like device. We label the device close to ground as the $n$-like device and the one close to $V_{DD}$ as the $p$-like device. For the devices to operate in a complementary circuit, one device has to switch from 'on' to 'off' ($p$-like) and one device from 'off' to 'on' ($n$-like). However, the devices have the same electronic state initially because they are both made from PANI. To obtain one device operating as $n$-like and the other as $p$-like (regions highlighted in yellow in **A** and **B**), we apply a $V_{shift} = -0.4$ V on the $n$-like device in the complementary circuit. At $V_{in} = 0$ V, the $n$-like device is at V = $-0.4$ V and it is in the 'off' state, whereas the $p$-like device is at V = 0 V and it is in the 'on' state.

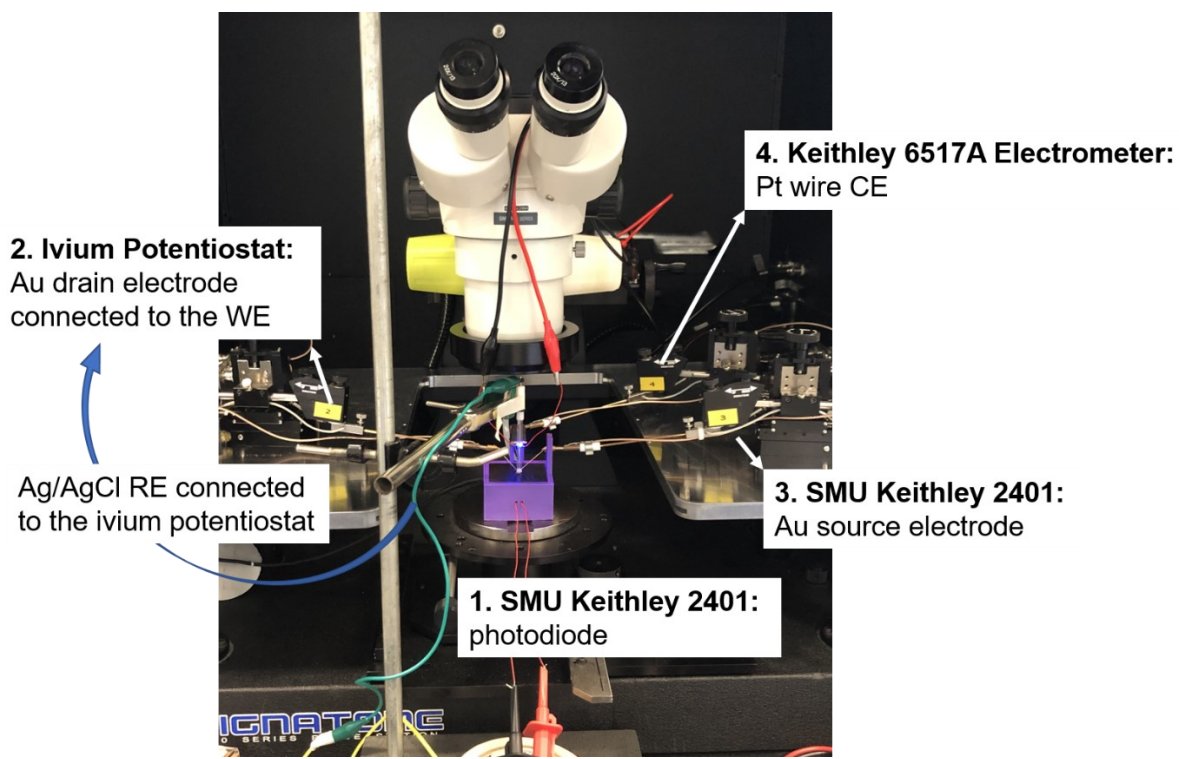

**Fig S5. In-house built apparatus for simultaneous electronic and optical characterisation of our OECT.** Ivium potentiostat (2) and Keithley 6517A electrometer (4) were used for measuring cyclic voltammetry. SMU Keithley 2401 (3) was used for applying $V_{DS}$ and measuring $I_D$. SMU Keithley 2401 (1) was used to measure the photodiode current. WE: working electrode; RE: reference electrode; CE: counter electrode.

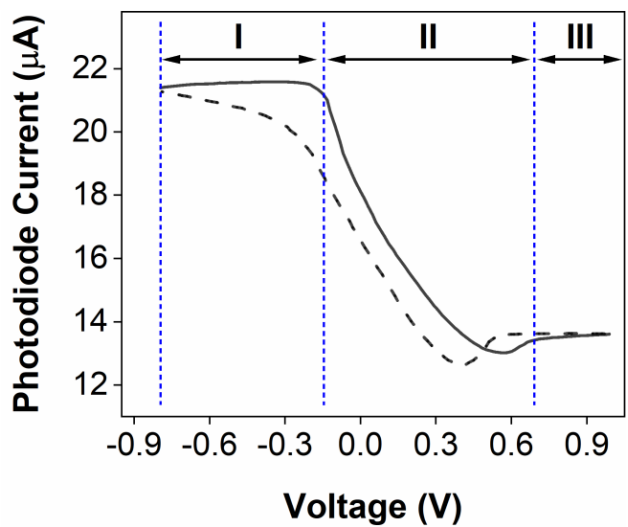

**Fig. S6. Changes in the photodiode current recorded as function of potential and related to the optical transmittance of the PANI film.** The measurement were performed under LED illumination using as a source LED with λ = 850 nm. Positive and negative sweeps are presented by solid and dashed lines respectively. The blue lines define the three distinct regions of potential: Region I (−0.8 to −0.1 V), Region II (−0.1 to +0.7 V), and Region III (+0.7 to +1.0 V).

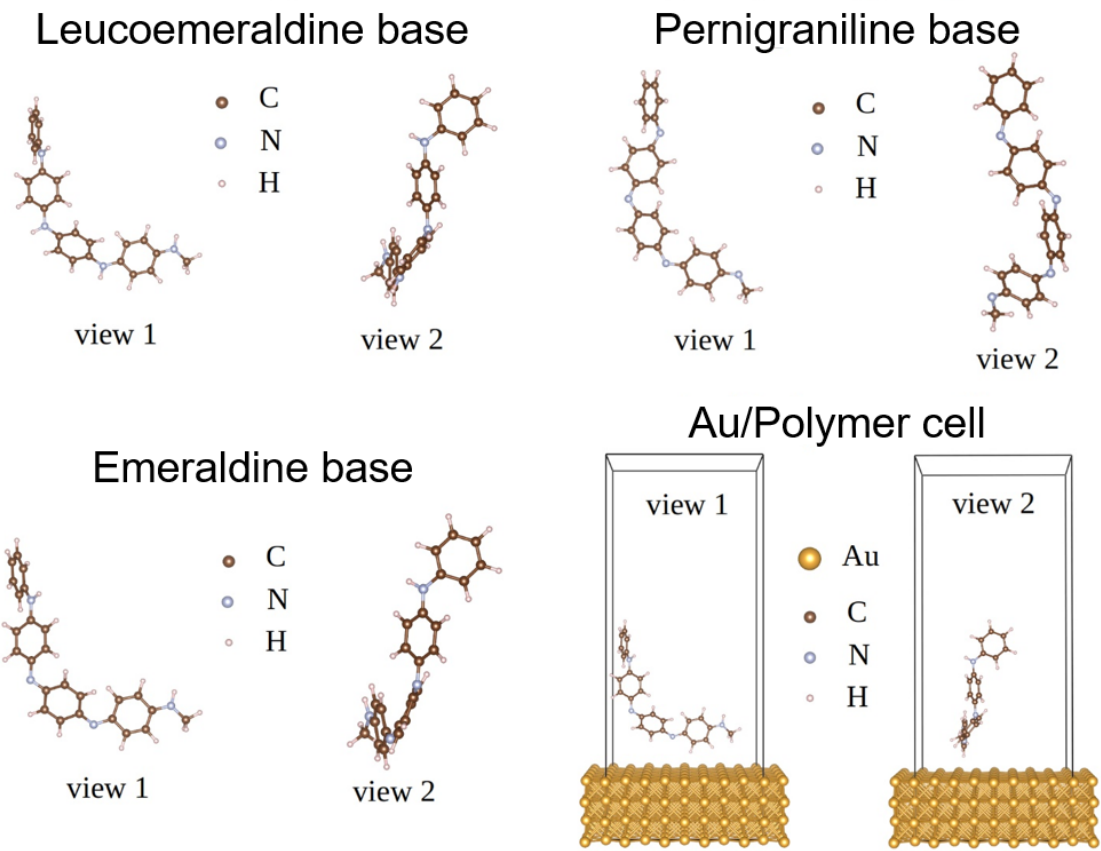

**Fig. S7. Details of the density functional theory simulations performed for Au substrate/PANI systems.** Different views of the polymers and simulation cell considered in the simulations are provided. The dominant interactions between the PANI and gold surface are hydrogen bonds of the Au⋯H-X type, as it is shown by the equilibrium geometry that renders lowest energy.

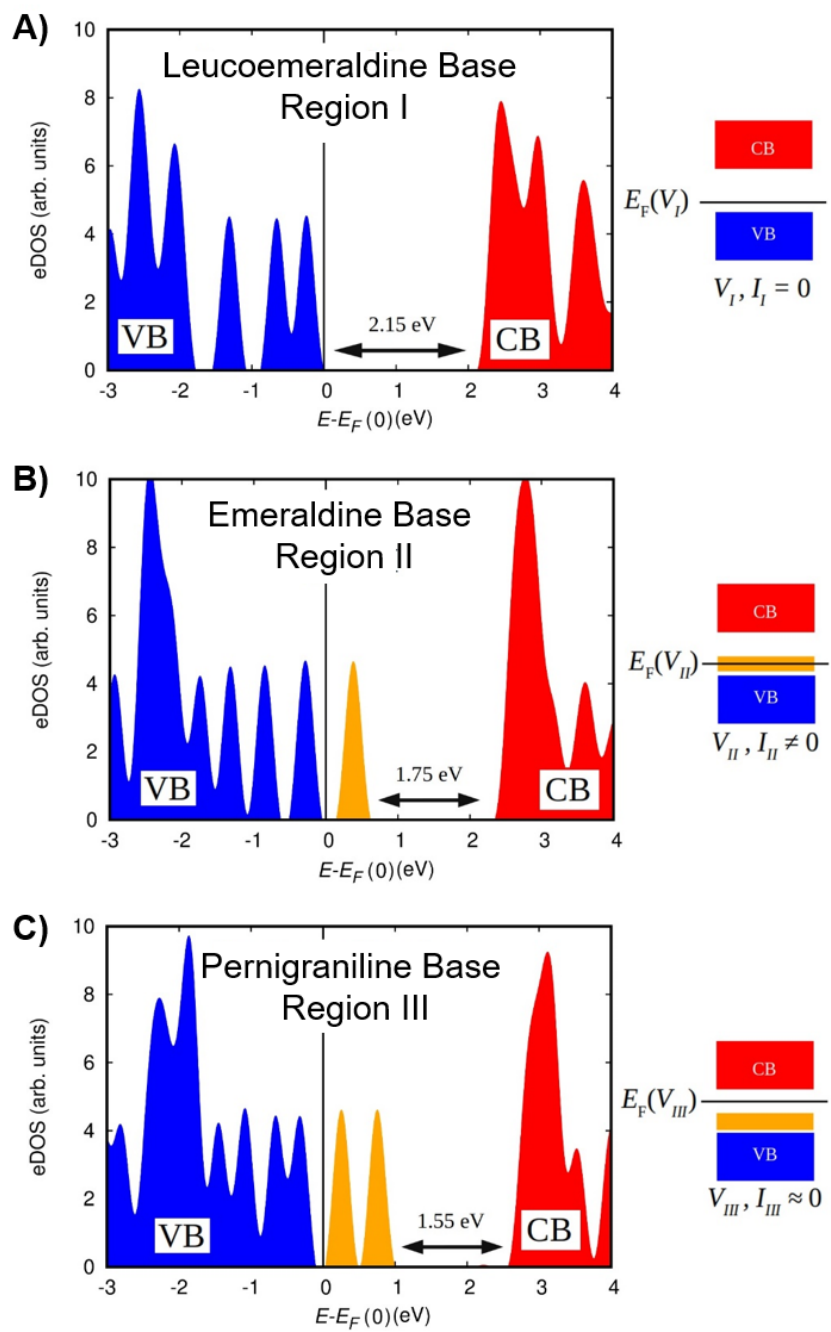

**Fig. S8. First-principles calculations of the electronic band structure of PANI.** The leucoemeraldine base is found to be insulator due to a wide band gap of 2.15 eV and position of the energy Fermi level just above the valence band edge. The emeraldine base is likely to be conductor under small electric bias due to the appearance of unoccupied electronic states at only ~0.1 eV above the valence band edge. The pernigraniline base is likely to be insulator under large enough electric bias due to the presence of a band gap of 1.55 eV just below the conduction band edge.

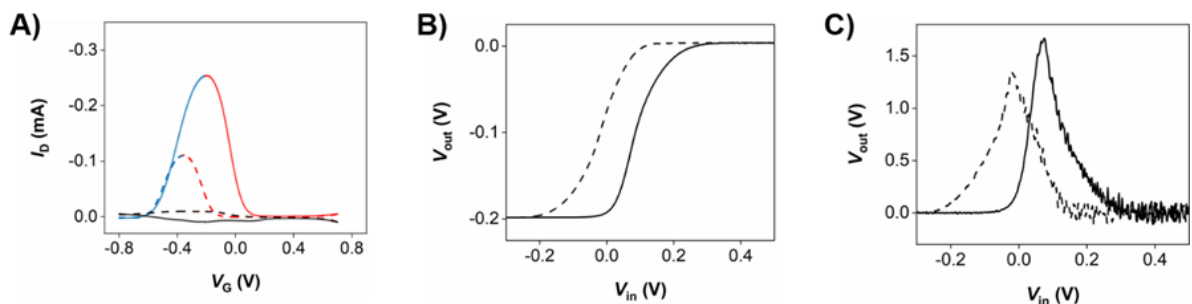

**Fig. S9. A flexible PANI-on-chitosan OECTs complementary circuit.** A) Drain current $I_D$ vs gate voltage $V_G$ at drain-source voltage $V_{DS} = -0.1$ V. Black lines represent the leakage current. B) Voltage-transfer characteristic for amplifier circuit made on flexible chitosan film operating with aqueous NaCl electrolyte. C) Corresponding voltage gain of the inverter.